\documentstyle[prl,aps,multicol]{revtex}
\input{epsf}
\begin{document}
\draft
\title{Dynamical localization simulated on a few qubits quantum computer}
\author{Giuliano Benenti$^{(a)}$, Giulio Casati$^{(a,b)}$, 
Simone Montangero$^{(a)}$, and Dima L. Shepelyansky$^{(c)}$}  
\address{$^{(a)}$International Center for the Study of Dynamical 
Systems, Universit\`a degli Studi dell'Insubria and} 
\address{Istituto Nazionale per la Fisica della Materia, 
Unit\`a di Como, Via Valleggio 11, 22100 Como, Italy}   
\address{$^{(b)}$Istituto Nazionale di Fisica Nucleare, 
Sezione di Milano, Via Celoria 16, 20133 Milano, Italy}   
\address{$^{(c)}$Laboratoire de Physique Quantique, UMR 5626 du CNRS,
Universit\'e Paul Sabatier, 31062 Toulouse Cedex 4, France}
\date{October 8, 2002}
\maketitle

\begin{abstract} 
We show that a quantum computer operating with a 
small number of qubits can simulate the dynamical 
localization of classical chaos in a system 
described by the quantum sawtooth map model. 
The dynamics of the system is computed efficiently 
up to a time $t\geq \ell$, and then the localization 
length $\ell$ can be obtained with accuracy $\nu$ by 
means of order $1/\nu^2$ computer runs, followed by 
coarse grained projective measurements on the computational 
basis. We also show that in the presence of static
imperfections a reliable computation of the localization 
length is possible without error correction up to an 
imperfection threshold which drops polynomially with 
the number of qubits. 
\end{abstract} 

\begin{multicols}{2}
\narrowtext

\section{Introduction} 

Recent experimental progress in nuclear magnetic 
resonance (NMR)-based quantum 
processors allowed the demonstration of quantum 
algorithms \cite{chuang}, including Grover's algorithm 
\cite{groverexp} and quantum Fourier transform 
\cite{qftexp}.
More recently, it has been possible to implement 
the simplest instance of Shor's algorithm, 
namely the factorization of 15, using 7 qubits 
and a sequence of about 300 spin-selective 
radio-frequency pulses \cite{shorexp}. 
In parallel, thanks to the development of techniques 
for the manipulation of cold atoms in linear traps,  
the realization of up to 50 two-qubit control-not gates 
within the relevant decoherence time scale is currently 
becoming possible\cite{blatt}. 
Solid state realizations are also under way in 
several experimental groups working with various 
solid state devices. In particular, it has been 
demonstrated that a superconducting tunnel junction 
circuit can behave as an artificial spin 1/2 atom, 
whose evolution can be controlled by applying microwave 
pulses. The quality factor of quantum coherence is 
sufficiently high to envisage the realization of 
two-qubit gates based on capacitively coupled circuits of 
this type \cite{saclay}. 

In this context, it is of primary importance to 
find efficient quantum algorithms, that could 
be usefully simulated with a small number of 
qubits. Such algorithms would naturally become 
the ideal software for demonstrative experiments 
in the coming generation of quantum processors.  
Dynamical models represent a natural testing 
ground for quantum information processors. 
The algorithm for the quantum baker's map \cite{schack} 
has been recently implemented on a three qubit NMR quantum 
processor \cite{bakerexp}. These experiments tested the 
sensitivity to perturbations, in a system which is 
characterized by chaotic unpredictable dynamics in 
the classical limit. 

In this paper we show that quantum computers can simulate 
efficiently the quantum localization of classical chaos. 
Dynamical localization is one of the most interesting
phenomena that characterize the quantum behavior of
classically chaotic systems: Quantum interference effects 
suppress chaotic diffusion in momentum, leading to
exponentially localized wave functions. This phenomenon 
was first 
found and studied  in the quantum kicked rotator model 
\cite{krot} and has profound analogies with Anderson 
localization of electronic transport in disordered materials 
\cite{fishman}.
Dynamical localization has 
been observed
experimentally in the microwave ionization of Rydberg 
atoms \cite{koch} and is now actively studied
in experiments with cold atoms \cite{raizen}.

In this paper, we study dynamical localization 
for the quantum sawtooth map, using the 
algorithm developed in Ref. \cite{saw1}. 
This algorithm has some specific advantages with 
respect to similar algorithms for the simulation 
of other dynamical systems, for instance the kicked 
rotator \cite{krotalg}. 
There are no extra work space qubits, namely 
all the qubits are used to simulate the  
dynamics of the system. This implies that 
less than 40 qubits would be sufficient to 
make simulations inaccessible to present day 
supercomputers. We note that this figure has 
to be compared with the more than 1000 qubits 
required to the Shor's algorithm to outperform 
classical computations. We will also show 
that in this model dynamical localization 
could be observed already with 6 qubits. 

The paper is organized as follows. In Section II
we describe the sawtooth map model, focusing 
on the regime of dynamical localization. 
In Section III we show that a quantum computer 
operating with few qubits can indeed perform 
simulations of dynamical localization. 
In Section IV we discuss how to extract 
information (the localization length) from 
the quantum computer wave function. In Section 
V we study the stability of those computations 
in the presence of static imperfections in the 
quantum computer hardware. In Section VI we 
discuss the transition to quantum chaos, induced 
by static imperfections, in the quasienergy 
spectral statistics. In Section VII we present our
conclusions. 

\section{The model}

The quantum sawtooth map is the quantized version 
of the classical sawtooth map, which is given by
\begin{equation}
\overline{n}={n}+k(\theta-\pi),
\quad
\overline{\theta}=\theta+T\overline{n},
\label{clmap}
\end{equation}
where $(n,\theta)$ are conjugated action-angle variables
($0\le \theta <2\pi$), and the over bars denote the 
variables after one map iteration. 
Introducing the rescaled momentum 
variable $p=Tn$, one can see that the classical dynamics 
depends only on the single parameter $K=kT$. 
The map (\ref{clmap}) 
can be studied on the cylinder ($p\in (-\infty,+\infty)$), 
which can also be closed to form a torus of 
length $2\pi L$, where $L$ is an integer.
For $K>0$, the motion is completely chaotic and exhibits 
normal diffusion: 
$<(\Delta p)^2> \approx D(K) t$, where $t$ is the discrete time
measured in units of map iterations and the average $<\cdots>$
is performed over an ensemble of particles with initial momentum
$p_0$ and random phases $0\leq \theta <2\pi$.
For $K>1$ the diffusion coefficient is well approximated by
the random phase approximation, $D(K)\approx (\pi^2/3) K^2$. 

The quantum evolution in one map iteration is described
by a unitary operator $\hat{U}$ (called Floquet operator) 
acting on the wave function $\psi$:
\begin{equation}
\overline{\psi}=\hat{U} \psi =
e^{-iT\hat{n}^2/2}
e^{ik(\hat{\theta}-\pi)^2/2}\psi,  
\label{qumap}
\end{equation}
where $\hat{n}=-i\partial/\partial\theta$ 
and $\psi(\theta + 2\pi) = \psi(\theta)$ (we set $\hbar=1$).
The classical limit corresponds to $k\to \infty$, 
$T\to 0$, and $K=kT=\hbox{const}$. 
In Refs. \cite{saw1,saw2,saw3} we studied the map 
(\ref{qumap}) in the semiclassical regime. 
This is possible by increasing the number of 
qubits $n_q=\log_2 N$ ($N$ is the total number of 
levels), with $T=2\pi L/N$, $K=\hbox{const}$. 
In this way, the number of levels inside the 
``unit cell'' $-\pi\leq p < \pi$ ($L=1$) 
grows exponentially with the number of qubits
($-N/2\leq n <N/2$), and the effective Planck constant 
$\hbar_{\rm eff}\sim\hbar/k\sim 1/N\to 0$ when 
$N\to \infty$. 

Differently from previous studies, 
in this paper we study the map (\ref{qumap}) in 
the deep quantum regime of dynamical localization. 
For this purpose, we keep $k,K$ constant. Thus the 
effective Planck constant is fixed and the number 
of cells $L$ grows exponentially with the number 
of qubits ($L=TN/2\pi$). 
In this case, one studies the 
quantum sawtooth map on the cylinder 
($n\in (-\infty,+\infty)$), which is cut-off 
to a finite number of cells due to the finite 
quantum (or classical) computer memory.
We stress again that, since in a quantum computer 
the memory capabilities grow exponentially with 
the number of qubits, already with less than 
40 qubits one could make simulations of 
systems inaccessible for today's supercomputers. 
Similar to other models of quantum chaos \cite{krot},
the quantum interference in the sawtooth map
leads to suppression of classical chaotic diffusion 
after a break time 
\begin{equation} 
t^\star\approx D_n\approx (\pi^2/3) k^2,
\label{loc1}
\end{equation}
where $D_n$ is the classical 
diffusion coefficient, measured in number of 
levels ($<(\Delta n)^2> \approx D_n t$). 
For $t > t^\star$ only $\Delta n \sim D_n$
levels are populated and the localization length $\ell \sim \Delta n$
for the average probability distribution
is approximately equal \cite{ds1987}:
\begin{equation}
\ell\approx D_n
\label{loc2}
\end{equation} 
Thus the quantum localization can be detected if $\ell$
is smaller than the system size $N$. 

In the following we consider $K=\sqrt{2}$, 
two values of $k$, $k=\sqrt{3}$ and $k=2$, 
and $6\leq n_q \leq 21$, so that the 
above analytical estimate gives 
$\ell(k=\sqrt{3})\approx 10$ and 
$\ell(k=2)\approx 13<N$.  
We assume that at $t=0$ the system is in a momentum 
eigenstate, $\hat{\psi}(n)=\delta_{n n_0}$. 
Since this is a quantum register state, it can be 
obtained in $O(n_q)$ one-qubit operations starting 
from the fiducial state (``ground state'') $\overline{n}$
of the quantum computer hardware. 

\section{Simulation of dynamical localization}

An exponentially efficient quantum algorithm 
for the simulation of 
the map (\ref{qumap}) was described in Refs. 
\cite{saw1,saw3}. It is based on the forward/backward 
quantum Fourier transform \cite{chuang} between 
momentum and coordinate bases: On the whole, it 
requires $n_g=3n_q^2+n_q$ gates per map iteration 
($3 n_q^2-n_q$ controlled-phase shifts 
and $2n_q$ Hadamard gates). 
This number has to be compared with the 
$O(N\log N)$ operations required to a classical 
computer to simulate one map iteration by means 
of the fast Fourier transform. 

In Fig. 1, we show that, using our quantum algorithm,  
exponential localization can be clearly seen already 
with $n_q=6$ qubits. 
After the break time $t^\star$, the probability 
distribution over the momentum eigenbasis decays 
exponentially, 
\begin{equation} 
W_n=|\hat{\psi}(n)|^2\approx 
\frac{1}{\ell}\exp\left(-\frac{2 |n-n_0|}{\ell}\right),
\label{expdecay}
\end{equation} 
with $n_0=0$ the initial momentum value. 
Here the localization length is  
$\ell\approx 12$, and classical diffusion 
is suppressed after a break time $t^\star\approx \ell$, 
in agreement with the estimates (\ref{loc1})-(\ref{loc2}). 
This requires a number 
$N_g \approx 3 n_q^2 \ell \sim 10^3$
of one- or two-qubit quantum gates.  
The full curve of Fig. 1 shows that an exponentially 
localized distribution indeed appears at $t\approx t^\star$. 
Such a distribution is frozen in time, apart from 
quantum fluctuations, which we partially smooth out by 
averaging over a few map steps. The localization
can be seen by the 
comparison of the probability distributions taken  
immediately after $t^\star$ (full curve in Fig. 1) 
and at a much larger time $t=300\approx 25 t^\star$ 
(dashed curve in the same figure). 

We also note that the asymptotic tails of the wave functions 
decay as a power law(see Fig. 2), 
\begin{equation}
W_n\propto \frac{1}{|n-n_0|^4}. 
\label{power}
\end{equation}
This happens due to the 
discontinuity in the kicking force of Eq. (\ref{clmap}), 
$f(\theta)=k(\theta-\pi)$, when the angle variable 
$\theta=0$. For that reason the matrix elements of 
the one period evolution operator $\hat{U}$ for 
the quantum map (\ref{qumap}) decay as a power law
in the momentum eigenbasis: 
$U_{nm}=\langle n|\hat{U}|m \rangle 
\sim 1/|n-m|^\alpha$, with $\alpha=2$. 
This case was investigated for random matrices, 
where it was shown that eigenfunctions are also 
algebraically localized with the same exponent 
$\alpha$ \cite{brm}. 
We also note that dynamical localization in 
discontinuous systems was studied in 
\cite{fausto}. 
Since the localization picture is not very 
sensitive to the behavior of the tails of the wave 
function, a rough estimate of the crossover 
between the exponential decay (\ref{expdecay}) and 
the power law decay (\ref{power}) is given by their 
crossing point,   
\begin{equation} 
n_c \sim \frac{3}{2} \ell \log \ell, \quad
W_n(n_c)\sim \frac{1}{\ell^4\log\ell}. 
\end{equation} 
This implies that by increasing $\ell$ the exponential 
localization is pushed to larger momentum windows and 
down to smaller probabilities. 

\begin{figure}
\centerline{\epsfxsize=8.cm\epsffile{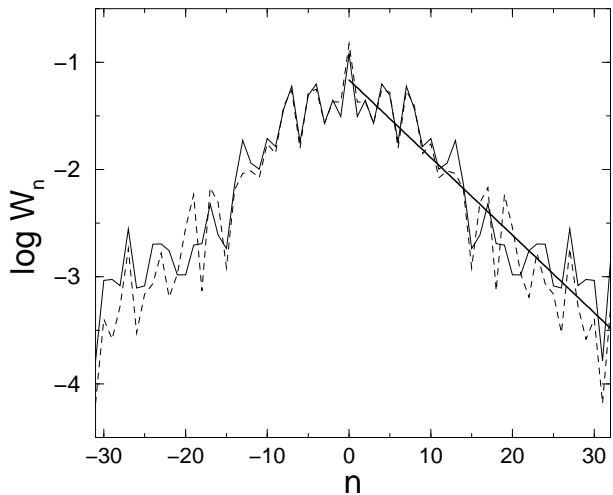}}
\caption{Exact quantum computation of probability distribution over the 
momentum basis with $n_q=6$ qubits for $k=\sqrt{3}$ and initial 
momentum $n_0=0$; 
the average is taken in the intervals $10\leq t \leq 20$
(full curve) and $290\leq t \leq 300$ 
(dashed curve). The straight line fit, 
$W_n\propto \exp(-2|n|/\ell)$, gives a 
localization length $\ell\approx 12$. Here and 
in the following figures, the logarithms are
decimal.}
\end{figure}

\section{Measurements}

We now discuss how it would be possible to 
extract information (the value of the localization 
length) from a quantum computer simulating the 
above described dynamics. 
The localization length can be measured by 
running the algorithm several times up to 
a time $t>t^\star$. Each run is followed 
by a standard projective measurement on 
the computational (momentum) basis. 
The outcomes of the measurements can be 
stored in histogram bins of width 
$\delta n \propto \ell$, 
and then the localization length can be 
extracted from a fit of the exponential decay 
of this coarse-grained distribution over the 
momentum basis. 
In this way the localization length can be 
obtained with accuracy $\nu$ after the order 
of $1/\nu^2$ computer runs. 
It is important to  note that
it is sufficient to perform a coarse grained measurement 
to generate a coarse grained distribution. 
This means that it will be sufficient to measure 
the most significant qubits, and ignore those 
that would give a measurement accuracy below 
the coarse graining $\delta n$. 
Thus, the number of runs and measurements is independent of $\ell$.
However, it is necessary to make about $t^\star \sim \ell$
map iterations to obtain the localized distribution 
(see Eqs. (\ref{loc1},\ref{loc2})).
This is true both 
for the present quantum algorithm and for 
classical computation. This implies that a 
classical computer needs $O(\ell^2\log \ell)$ 
operations to extract the localization 
length, while a quantum computer would 
require $O(\ell (\log \ell)^2)$ elementary gates
(classically one can use a basis size $N \sim \ell$
to detect localization). 
In this sense, for $\ell \sim N=2^{n_q}$ 
the quantum computer gives a 
square root speed up if both classical and quantum computers
perform $O(N)$ map iterations.
However, for a fixed number of iterations $t$
the quantum computation gives an exponential gain.
For $\ell \ll N$ such a gain can be very important
for  more complex 
physical models, in order to check if the 
system is truly localized \cite{harper}.

\section{Effects of static imperfections}

In order to study the effects of static imperfections
on the stability of the above described algorithm, 
we model the quantum computer hardware as a linear 
array of qubits with static imperfections, 
represented by  fluctuations in the individual 
qubit energies and residual short-range inter-qubit 
couplings \cite{GS}. 
The model is described by the following many-body Hamiltonian:  
\begin{equation}
\hat{H}_{\hbox{s}}=\sum_i (\Delta_0+\delta_i)\hat{\sigma}_i^z +
\sum_{i<j}J_{ij}\hat{\sigma}_i^x\hat{\sigma}_j^x,
\label{imperf}
\end{equation}
where the $\hat{\sigma}_i$i's are the Pauli matrices for the qubit 
$i$, and $\Delta_0$ is the average level spacing for one qubit. 
The second sum in (\ref{imperf}) runs over nearest-neighbor qubit 
pairs, zero boundary conditions are applied, 
and $\delta_i$, $J_{ij}$ are randomly and uniformly distributed
in the intervals $[-\delta/2,\delta/2]$ and $[-J,J]$, respectively.
We model the implementation of the above algorithm on this 
hardware architecture as a sequence of instantaneous and perfect 
one- and two-qubit gates, separated by a time interval $\tau_g$. 
Therefore we study numerically the evolution in time of the 
quantum computer wave function in the presence of the following 
many-body Hamiltonian: 
\begin{equation}
\hat H( \tau ) = \hat H_{\hbox{s}} + \hat H_{\hbox{g}}( \tau ), 
\label{himp} 
\end{equation}
where 
\begin{equation}
\hat H_{\hbox{g}}(\tau ) = \sum_k \delta (\tau  - k \tau_g ) \hat h_k.
\end{equation}
Here $\hat h_k$ realizes  the $k$-th elementary gate according to 
the sequence prescribed by the algorithm. 
We assume that the phase accumulation given by $\Delta_0$ 
is eliminated by standard spin echo techniques \cite{chuang}.
In this case, the remaining terms in the static Hamiltonian 
(\ref{imperf}) can be seen as residual terms after imperfect
spin echoes and give unwanted phase rotations and qubit 
couplings. 

The effect of static imperfections on 
the probability distribution over the momentum basis 
is shown in Fig. 2, for 
$k=2$, $n_q=11$, $t=100$, $J=0$, 
and different rescaled imperfection strengths
$\epsilon=\delta \tau_g$. For $\epsilon=10^{-4}$, 
the localization peak is reproduced with 
high fidelity, while the tails of the wave function 
are strongly enhanced. This is due to the fact 
that errors affecting the most significant 
qubits can induce a direct transfer of probability 
very far in the momentum basis \cite{song,saw1}. 
For $\epsilon=10^{-3}$ a measurement of the 
decay of the localization peak would overestimate 
the localization length by a factor two, 
while for $\epsilon=10^{-2}$ any trace of dynamical 
localization has been destroyed. 

\begin{figure}
\centerline{\epsfxsize=8.cm\epsffile{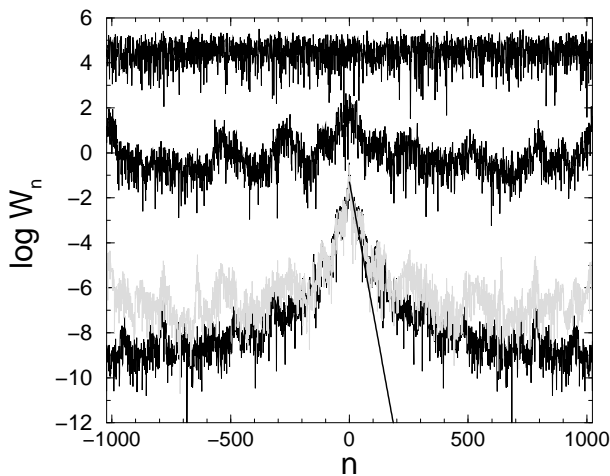}}
\caption{Probability distributions for $k=2$,  
$n_q=11$, $n_0=0$, $J=0$. From bottom to top:
$\epsilon=0$, $10^{-4}$ (gray line), 
$10^{-3}$ (shifted up by a factor $4$), 
and $10^{-2}$ (shifted up by a factor $8$).
The straight line gives a localization length 
$\ell\approx 15$.} 
\end{figure} 

In order to study in a more quantitative way the stability 
of quantum computation in the presence of static 
imperfections, we consider the following two quantities:
\newline
(i) the diffusion coefficient $D_n(\epsilon)$, 
obtained from the relation 
\begin{equation} 
<(\Delta n)^2>\approx D_n(\epsilon) t.
\end{equation} 
This is an important characteristic related 
to transport properties of the system; 
\newline 
(ii) the inverse participation ratio 
\begin{equation} 
\xi=\frac{1}{\sum_n W_n^2}; 
\end{equation} 
this quantity determines the number of 
basis states significantly populated by the wave 
function and gives an estimate of the localization 
length of the system. 
We stress that, differently from the previous 
quantity, $\xi$ is local in the localized regime, 
i.e. it is insensitive to the behavior of 
exponentially small tails. 

\begin{figure}
\centerline{\epsfxsize=8.cm\epsffile{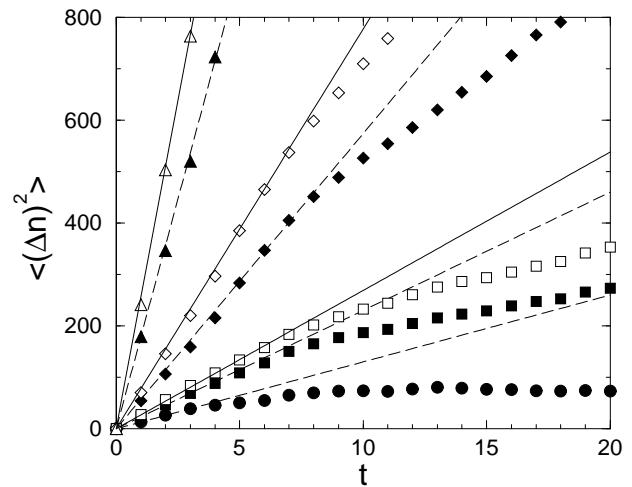}}
\caption{Dependence of the wave function second 
moment on time, for $k=2$, $n_q=11$, $J=0$ 
(full symbols) and $J=\delta$ (empty symbols) at 
$\epsilon=0$ (circles), $5\times 10^{-5}$ (squares),
$10^{-4}$ (diamonds), and $2\times 10^{-4}$ (triangles). 
The straight lines fits give the diffusion coefficient 
$D_n(\epsilon)$. 
The curves are averaged over $10$ disorder 
realizations and $10$ initial conditions 
$n_0\in [-5,5[$.}
\end{figure} 

In Fig. 3 we show $<(\Delta n)^2>$ as a function of 
time, for $n_q=11$ qubits, $J=0$, and different 
imperfection strengths $\epsilon$. By means of 
these curves we extract the diffusion coefficients 
$D_n(\epsilon)$ from linear fits extended to the 
first few map steps. In the same figure, we show 
that similar curves are obtained for $J=\delta$. 
The dependence of the inverse participation ratio 
on $t$ is shown in Fig. 4, again for $k=2$,
$n_q=11$. 
We note that, for 
imperfection strengths strong
enough to induce huge variations in the diffusion
coefficient ($D_n(\epsilon)\gg D_n(0)$), 
$\xi$ is only slightly modified ($\xi(\epsilon)
\approx \xi(0)$). 
Iterating the map (\ref{qumap}) long enough 
($t>\xi(\epsilon)$), we get the saturation 
value $\xi_\infty (\epsilon)$. 
This quantity increases with $\epsilon$ and one 
has complete delocalization when $\xi_\infty
(\epsilon)\sim N$ (this is evident for 
$\epsilon=5\times 10^-3$ in 
Fig. 4; in this case $\xi$ saturates 
after $t<100$ map iterations). 
Again we note that similar curves 
are obtained for $J=\delta$ (see Fig. 4). 

\begin{figure}
\centerline{\epsfxsize=8.cm\epsffile{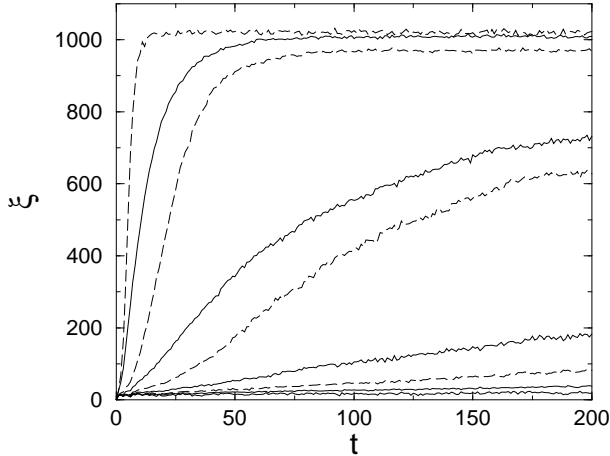}}
\caption{Time dependence of the inverse participation 
ratio $\xi$ for $k=2$, $n_q=11$ qubits, $J=0$ (solid lines) 
and $J=\delta$ (dashed lines). From bottom to top:
$\epsilon=0$, $5\times 10^{-4}$, $10^{-3}$,
$2\times 10^{-3}$, $5\times 10^{-3}$. The curves 
are averaged as in Fig. 3.}
\end{figure} 

In Fig. 5 we plot the dependence of diffusion 
coefficient $D_n$ on $\epsilon$ for different 
$n_q$ values. From each curve 
we extract the critical imperfection strength 
$\epsilon_D(n_q)$ corresponding to 
doubling of the diffusion 
coefficient, $D_n(\epsilon_D)=2D_n(0)$. 
The data of Fig. 7 show that $\epsilon_D$ drops 
exponentially with $n_q$. This 
result is similar to the one found in \cite{song} 
for noisy gate errors and
can be explained by 
means of the following argument. 
Static imperfections can couple states very far 
in momentum space via a single spin flip. 
As a consequence, they create $n_q$ peaks 
\cite{song} with probability $W_p\sim 
\epsilon_{\rm eff}^2 t$ in each peak. Here 
$\epsilon_{\rm eff}\sim \delta n_q^2 \tau_g =
\epsilon n_q^2$ is the effective perturbation 
strength, with $n_q^2 \tau_g$ time between 
Hadamard gates acting on a given qubit. These 
gates transfer the accumulated phase error 
$\epsilon_{\rm eff}$ into amplitude errors. 
Integrating the contribution of each peak, 
one gets
\begin{equation} 
<(\Delta n)^2> \sim W_p N^2 \sim 
\overline{D}_\epsilon t, 
\end{equation} 
with 
\begin{equation} 
\overline{D}_\epsilon\sim \epsilon^2 n_q^4 N^2. 
\label{deps}
\end{equation} 
One can estimate the critical value $\epsilon_D$ 
to double the exact ($\epsilon=0$) 
diffusion coefficient 
from $\overline{D}_\epsilon=D_n(\epsilon=0)$, giving 
\begin{equation} 
\epsilon_D\sim \frac{\sqrt{D_n(0)}}{n_q^2 N}, 
\end{equation} 
in good agreement with the data of Fig. 7. 

\begin{figure}
\centerline{\epsfxsize=8.cm\epsffile{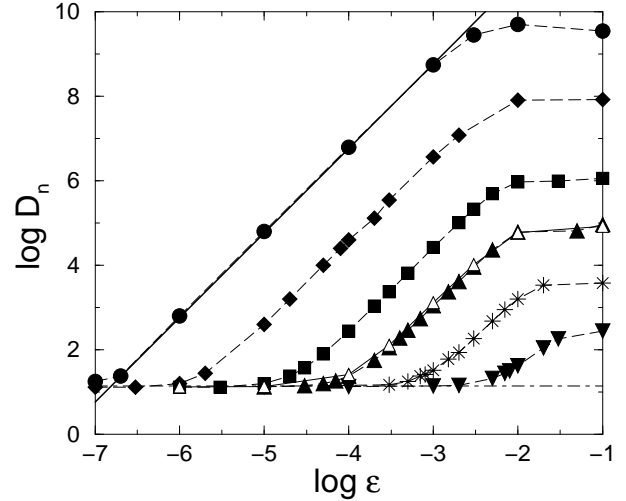}}
\caption{Dependence of the diffusion coefficient 
$D_n$ on the imperfection strength $\epsilon$ 
for $k=2$, $J=0$ (full symbols), 
$n_q=6$ (triangles down), $8$ (stars),
$10$ (triangles up), $12$ (squares), 
$15$ (diamonds), $18$ (circles), 
and for $J=\delta$, $n_q=10$ (empty triangles). 
The straight lines show the theoretical 
dependence $D\propto \epsilon^2$ (full line, 
see Eq. (\ref{deps}))
and the result without imperfections, 
$D(\epsilon=0)\approx 16$ (chain line).} 
\end{figure} 

\begin{figure}
\centerline{\epsfxsize=8.cm\epsffile{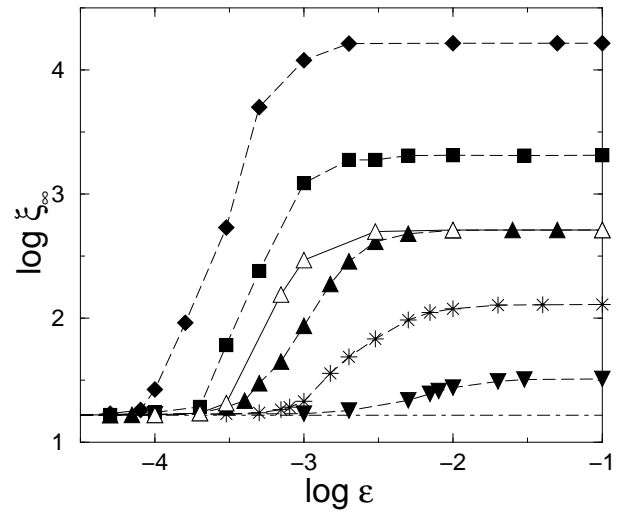}}
\caption{Dependence of the saturation value 
$\xi_\infty$ of the inverse participation ratio 
on the imperfection strength $\epsilon$, 
with same parameter values and same meaning of 
symbols as in the previous figure.  
The straight line shows the result without 
imperfections, $\xi_\infty(\epsilon=0)
\approx 14$.} 
\end{figure} 

In Fig. 6 we show the dependence of the inverse 
participation ratio $\xi(\epsilon)$ on $\epsilon$ 
for different number of qubits $n_q$. From each curve we extract 
two critical imperfection strengths: 
\newline 
(i) $\epsilon_{\xi E}$, to get an inverse participation 
ratio equal to a given fraction of the full Hilbert 
space, for example $\xi=N/4$. This threshold is 
significant of the transition to ergodic 
completely delocalized wave functions. 
\newline 
(ii) $\epsilon_\xi$, to double the exact inverse 
participation ratio. We stress that 
this quantity gives a rough 
estimate of the imperfection threshold for 
reliable quantum computation of localization 
in the absence of error correction. 

The dependence of $\epsilon_{\xi E}$ and $\epsilon_\xi$ 
on $n_q$ are shown in Fig. 7. These quantities drop 
polynomially with $n_q$, in sharp contrast with 
the exponential drop of $\epsilon_D$. This algebraic 
threshold can be understood as follows. 
The eigenstates of the
unperturbed ($\epsilon=0$, J=0) Floquet operator 
$\hat{U}$ in (\ref{qumap}) can be written as 
\begin{equation}
\phi_\alpha^{(0)}= \sum_{m=1}^N c_\alpha^{(m)} u_m, 
\end{equation} 
where $u_m$ are the quantum register (momentum) states. 
In the localized regime, $c_\alpha^{(m)}$'s are 
randomly fluctuating inside the localization 
domain of size $\ell$, and exponentially small 
outside it. Wave function normalization imposes  
$|c_\alpha^{(m)}| \sim 1/\sqrt{\ell}$.
Due to exponential localization, static imperfections 
couple significantly the unperturbed eigenfunctions only 
when their localization domains overlap. We estimate 
in this case the transition matrix elements according 
to perturbation theory. For $J=0$, they have a typical 
value 
\begin{eqnarray}
V_{\rm typ} \sim |\langle\phi_\beta^{(0)}|
\sum_{i=1}^{n_q}\delta_i\hat{\sigma}_i^z \tau_g n_g
| \phi_\alpha^{(0)}\rangle| \nonumber 
\\ \sim  
\tau_g n_q^{2} |\sum_{m,n=1}^{\ell} 
c_\alpha^{(m)} c_\beta^{(n)\star}
\sum_{i=1}^{n_q} \delta_i 
\langle u_n | \hat{\sigma}_i^z | u_m \rangle | 
\label{vtyp}\\
\sim \epsilon n_q^{5/2} |\sum_{m=1}^{\ell} 
c_\alpha^{(m)} c_\beta^{(m)\star}\eta^{(m)}|
\sim \epsilon n_q^{5/2}\ell^{-1/2}. \nonumber
\end{eqnarray}
In this expression, the typical phase error  
is $\delta \sqrt{n_q} \eta^{(m)}$ 
(sum of $n_q$ random detunings $\delta_i$'s),  
with $\eta^{(m)}$ random sign, and 
$\tau_g n_g \sim \tau_g n_q^2$ is the 
time used by the quantum computer to simulate one map step.
The last estimate in (\ref{vtyp}) results from the sum of 
order $\ell$ terms of amplitude 
$|c_\alpha^{(m)} c_\beta^{(m)\star}| \sim 1/\ell$ 
and random phases.
Since the spacing between significantly coupled 
quasi-energy eigenstates is
$\Delta E \sim 1/\ell$, the threshold for the 
breaking of perturbation theory can be estimated as
\begin{equation}
V_{\rm typ}/\Delta E \sim \epsilon_{\xi E}
n_q^{5/2}\sqrt{\ell} \sim 1.
\label{chaosth}
\end{equation} 
The analytical result 
\begin{equation} 
\epsilon_{\xi E}\sim \frac{1}{n_q^{5/2}\sqrt{\ell}}
\label{ergest} 
\end{equation} 
is confirmed by the numerical data of Fig. 7.
For the case $J=\delta$, the threshold $\epsilon_{\xi E}$
is reduced (see Fig. 4) since residual inter-qubit 
interactions introduce further couplings between 
Floquet eigenstates. However, an estimate similar 
to (\ref{vtyp}), which does not modify the functional 
dependence (\ref{chaosth}), can be derived. 
We stress the striking difference between this 
polynomial scaling and the exponential scaling 
for the mixing of unperturbed eigenstates obtained 
in the ergodic regime (in which $\ell\sim N$) and 
in the more general quasi-integrable regime 
\cite{saw2,saw3}. 
We also note that the different sensitivity of local 
and non local quantities was pointed out in 
\cite{song}. However, the authors of Ref. \cite{song} 
considered the effect of noisy gates, while we 
consider internal static imperfections. 

\begin{figure}
\centerline{\epsfxsize=8.cm\epsffile{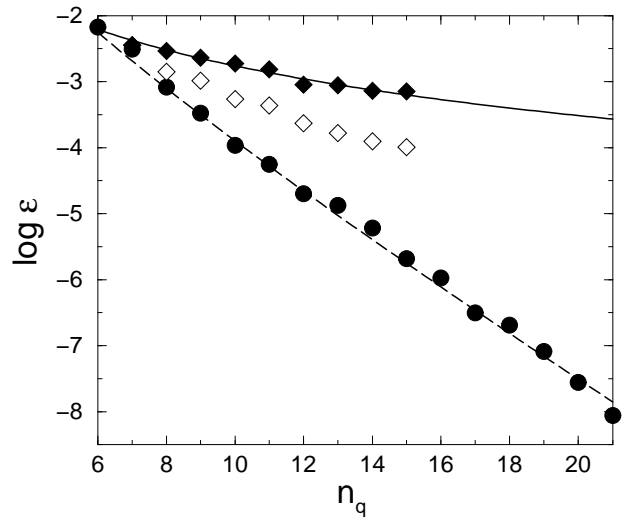}}
\caption{Dependence of the critical imperfection 
strengths on the number of 
qubits for $k=2$, $J=0$: 
thresholds $\epsilon_D$ (circles), 
$\epsilon_{\xi E}$ (full diamonds),
and $\epsilon_\xi$ (empty diamonds).  
The full line gives the theoretical dependence 
$\epsilon_{\xi E}=A n_q^{-5/2}$, with the 
fitting constant $A\approx 0.5$.
The dashed lines gives 
$\epsilon_D = B \sqrt{D(\epsilon=0)} 
2^{-n_q} n_q^{-2}$, with the 
fitting constant $B\approx 3.6$.}
\end{figure} 

\section{Spectral statistics} 

In this Section, we show that spectral statistics is an 
ideally suited tool to detect the destruction of 
localization by static imperfections. 
We study the spectral statistics of 
the Floquet operator for a quantum 
computer running the quantum sawtooth map 
algorithm in the presence of static imperfections, 
\begin{equation} 
\hat{U}_\epsilon=
\exp \left(-i\int_0^{\tau_g n_g} 
d \tau \hat{H} (\tau ) \right),
\end{equation} 
where $H(\tau)$ is the Hamiltonian (\ref{himp}) and 
$n_g$ the number of gates per map iteration.
We construct numerically the Floquet operator in 
the computational (momentum) basis, using the 
fact that a single map iteration of each quantum 
register state gives a column in the matrix representation 
of this operator. Then we diagonalize the Floquet 
matrix and get the so-called quasienergy eigenvalues
$\lambda_\alpha^{(\epsilon)}$ and eigenvectors 
$\phi_\alpha^{(\epsilon)}$,
\begin{equation} 
\hat{U}_\epsilon 
\phi_\alpha^{(\epsilon)}=\exp(i \lambda_\alpha^{(\epsilon)})
\phi_\alpha^{(\epsilon)}.
\end{equation} 

A convenient way to characterize the spectral properties 
of the system is to study the level spacing statistics 
$P(s)$, where $P(s)ds$ gives the probability to find two 
adjacent levels (quasienergies) whose energy difference,
normalized to the average level spacing, belongs to the 
interval $[s,s+ds]$
(see e.g. \cite{bohigas,guhr}).
In the localized regime, Floquet eigenvectors with 
very close eigenvalues may lay so far apart that 
their overlap is negligible. As a consequence, 
eigenvalues are uncorrelated, that is 
their spectral statistic is given by the Poisson 
distribution, 
\begin{equation} 
P_P(s)=\exp(-s). 
\label{poisson}
\end{equation} 
On the contrary, in the delocalized regime 
wave functions are ergodic, and their overlap 
gives a significant coupling matrix element 
between states nearby in energy. In this case 
the spectral 
statistics $P(s)$ follows the Wigner-Dyson 
distribution,  
\begin{equation} 
P_{WD}(s)=\frac{32 s^2}{\pi^2}\exp\left(
-\frac{4 s^2}{\pi} \right), 
\label{WD}
\end{equation} 
typical of random matrices in the absence of 
time reversal symmetry \cite{bohigas,guhr} 
(static imperfections break this symmetry). 
In Fig. 8 we show that static imperfections 
indeed induce a crossover 
from the localized regime
with the  
Poisson statistics to quantum chaos characterized by 
Wigner-Dyson statistics. 
We have also studied this crossover as a function 
of the number of qubits (data not shown): the threshold 
$\epsilon_c(n_q)$ for the emergence of quantum chaos is 
consistent with the scaling $\epsilon_c(n_q)\propto 
n_q^{-5/2}$, in agreement with the threshold 
(\ref{ergest}) 
obtained for the mixing of unperturbed 
eigenfunctions. 

\begin{figure}
\centerline{\epsfxsize=8.cm\epsffile{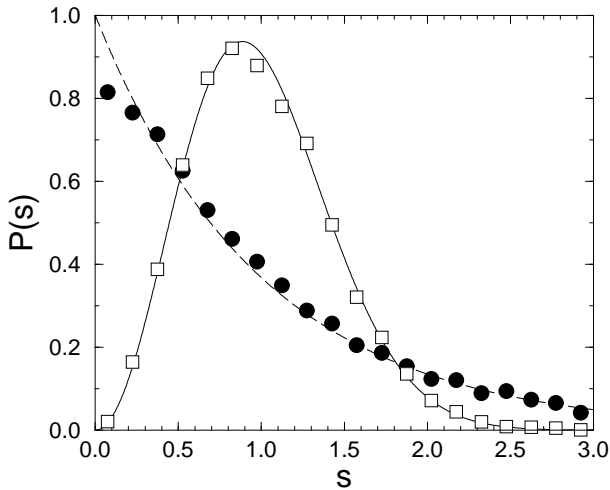}}
\caption{Level spacing statistics 
for $k=2$, $n_q=11$, 
$J=0$, $\epsilon=10^{-5}$ (circles) and 
$\epsilon=2.6\times 10^{-3}$ (squares). 
The dashed and full curves give the Poisson 
(\ref{poisson}) and Wigner-Dyson distribution
(\ref{WD}), respectively.
In order to reduce statistical fluctuations, 
data are averaged over $N_D=5$ random 
realizations of $\delta_i$'s, so that 
the total number of spacings is 
$N_D N \approx 10^4$.}  
\end{figure}

\section{Conclusions} 

In summary, we have shown that a quantum computer 
operating with a small number of qubits can 
simulate efficiently quantum localization effects.
The evaluation of the localization length $\ell$ 
with accuracy $\nu$ requires a number of computer 
runs of order $1/\nu^2$, followed by a projective 
measurement in the computational (momentum) basis. 
We stress that, in the presence of static imperfections, 
a reliable computation of the localization length 
is possible even without quantum error correction, 
up to an imperfection strength threshold which 
drops only algebraically with the number of qubits. 
We also stress that localization is a purely 
quantum phenomenon, which is quite fragile 
in the presence of noise \cite{ott,song}. 
Therefore we believe that the simulation of 
the physics of localization can be an interesting 
testing ground for the coming generation of quantum 
processors operating in the presence of decoherence
and static imperfections.

This research was supported in part by the EC RTN contract 
HPRN-CT-2000-0156, the NSA and ARDA under 
ARO contracts No. DAAD19-01-1-0553
and No. DAAD19-02-1-0086, 
the project EDIQIP of the IST-FET programme of the EC
and the
PRIN-2000 ``Chaos and localization in classical and
quantum mechanics''.

\end{multicols}

\end{document}